\documentclass[12pt]{article}

\usepackage{amsmath,amsfonts,amssymb}
\usepackage{epsfig,graphicx,bm}
\usepackage{epstopdf}
\begin{document}

\begin{titlepage}

\begin{center}

\vskip 0.4 cm

\begin{center}
{\Large  \bf Note About Unstable D-Brane with Dynamical Tension }
\end{center}

\vskip 1cm

\vspace{1em} J. Kluso\v{n},$^{a,}$
\footnote{Email address: klu@physics.muni.cz , }\\
\vspace{1em}$^a$\textit{Department of Theoretical Physics and
Astrophysics, Faculty of Science,\\
Masaryk University, Kotl\'a\v{r}sk\'a 2, 611 37, Brno, Czech Republic}
\vskip 0.8cm

\end{center}

\begin{abstract}
We propose an action for unstable Dp-brane with dynamical tension. We show that
the equations of motion are equivalent to the equations of motion derived from
DBI and WZ actions for non-BPS Dp-brane. We also find Hamiltonian formulation of this action
and analyze properties of the solutions corresponding to the tachyon vacuum and zero
tension solution.

\end{abstract}

\end{titlepage}

\bigskip

\newpage

\def\det{\mathrm{det}}
\def\tR{\tilde{R}}
\def\th{\tilde{h}}
\def\bZ{\mathbf{Z}}
\def\bp{\mathbf{p}}
\def\hrho{\hat{\rho}}
\def\tf{\tilde{f}}
\def\bPi{\mathbf{\Pi}}
\def\tmG{\tilde{\mG}}
\def\tJ{\tilde{J}}
\def\hx{\hat{x}}
\def\tr{\mathrm{Tr}}
\def\hC{\hat{C}}
\def\ttau{\tilde{\tau}}
\def\tmH{\tilde{\mH}}
\def\tbA{\tilde{\bA}}
\def\tsigma{\tilde{\sigma}}
\def\tJ{\tilde{J}}
\def\hT{\hat{T}}
\def\str{\mathrm{Str}}
\def\bp{\mathbf{p}}
\def\Pf{\mathrm{Pf}}
\def\mF{\mathcal{F}}
\def\tx{\tilde{x}}
\def\tF{\tilde{F}}
\def\I{\mathbf{i}}
\def\mM{\mathcal{M}}
\def\IT{\I_{\Phi,\Phi',T}}
\def\hf{\hat{f}}
\def\bx{\mathbf{x}}
\def \cit{\IT^{\dag}}
\def\tkappa{\tilde{\kappa}}
\def \cdt{\overline{\tilde{D}T}}
\def \dt{\tilde{D}T}
\def\bra #1{\left<#1\right|}
\def\ket #1{\left|#1\right>}
\def\vac #1{\left<\left<#1\right>\right>}
\def\pb  #1{\left\{#1\right\}}
\def \uw #1{(w^{#1})}
\def\mC{\mathcal{C}}
\def \dw #1{(w_{#1})}
\newcommand{\thw}{\tilde{\hat{w}}}
\def\bm{\mathbf{m}}
\def\bB{\mathbf{B}}
\newcommand{\bA}{{\bf A}}
\newcommand{\bd}{{\bf d}}
\newcommand{\bD}{{\bf D}}
\newcommand{\bF}{{\bf F}}
\newcommand{\bN}{{\bf N}}
\newcommand{\hp}{\hat{p}}
\newcommand{\hq}{\hat{q}}
\newcommand{\hF}{\hat{F}}
\newcommand{\hG}{\hat{G}}
\newcommand{\hH}{\hat{H}}
\newcommand{\bK}{\mathbf{K}}
\newcommand{\hU}{\hat{U}}
\newcommand{\mH}{\mathcal{H}}
\newcommand{\mG}{\mathcal{G}}
\newcommand{\mA}{\mathcal{A}}
\newcommand{\mD}{\mathcal{D}}
\newcommand{\tpr}{t^{\prime}}
\newcommand{\bzg}{\overline{\zg}}
\newcommand{\of}{\overline{f}}
\newcommand{\ow}{\overline{w}}
\newcommand{\htheta}{\hat{\theta}}
\newcommand{\opartial}{\overline{\partial}}
\newcommand{\hd}{\hat{d}}
\newcommand{\halpha}{\hat{\alpha}}
\newcommand{\hbeta}{\hat{\beta}}
\newcommand{\hdelta}{\hat{\delta}}
\newcommand{\hgamma}{\hat{\gamma}}
\def\mK{\mathcal{K}}
\newcommand{\hlambda}{\hat{\lambda}}
\newcommand{\hw}{\hat{w}}
\newcommand{\hN}{\hat{N}}
\newcommand{\onabla}{\overline{\nabla}}
\newcommand{\hmu}{\hat{\mu}}
\newcommand{\hnu}{\hat{\nu}}
\newcommand{\ha}{\hat{a}}
\newcommand{\hb}{\hat{b}}
\def\hr{\hat{r}}
\newcommand{\hc}{\hat{c}}
\newcommand{\com}[1]{\left[#1\right]}
\newcommand{\oz}{\overline{z}}
\newcommand{\oJ}{\overline{J}}
\newcommand{\mL}{\mathcal{L}}
\newcommand{\oh}{\overline{h}}
\newcommand{\oT}{\overline{T}}
\newcommand{\oepsilon}{\overline{\epsilon}}
\newcommand{\tP}{\tilde{P}}
\newcommand{\hP}{\hat{P}}
\newcommand{\uc}{\underline{c}}
\newcommand{\ud}{\underline{d}}
\newcommand{\ue}{\underline{e}}
\newcommand{\uf}{\underline{f}}
\newcommand{\hpi}{\hat{\pi}}
\newcommand{\oZ}{\overline{Z}}
\newcommand{\tg}{\tilde{g}}
\def\hbA{\hat{\bA}}
\newcommand{\tK}{\tilde{K}}
\newcommand{\tG}{\tilde{G}}
\newcommand{\hg}{\hat{g}}
\newcommand{\htG}{\hat{\tilde{G}}}
\newcommand{\hX}{\hat{X}}
\newcommand{\hY}{\hat{Y}}
\newcommand{\hthteta}{\hat{\theta}}
\newcommand{\hB}{\hat{B}}
\newcommand{\tlambda}{\tilde{\lambda}}
\newcommand{\thlambda}{\tilde{\hat{\lambda}}}
\newcommand{\tw}{\tilde{w}}
\newcommand{\hJ}{\hat{J}}
\newcommand{\tPsi}{\tilde{\Psi}}
\newcommand{\cP}{{\cal P}}
\newcommand{\tOmega}{\tilde{\Omega}}
\newcommand{\homega}{\hat{\omega}}
\newcommand{\hupsilon}{\hat{\upsilon}}
\newcommand{\hUpsilon}{\hat{\Upsilon}}
\newcommand{\hOmega}{\hat{\Omega}}
\newcommand{\bJ}{\mathbf{J}}
\newcommand{\olambda}{\overline{\lambda}}
\newcommand{\uhlambda}{\underline{\hlambda}}
\newcommand{\uhw}{\underline{\hw}}
\def \lhw #1{(\hw^{#1})}
\def \dhw #1{(\hw_{#1})}
\newcommand{\bG}{\mathbf{G}}
\newcommand{\bhG}{\hat{\bG}}
\newcommand{\bH}{\mathbf{H}}
\newcommand{\bE}{\mathbf{E}}
\newcommand{\mJ}{\mathcal{J}}
\newcommand{\mY}{\mathcal{Y}}
\newcommand{\mZ}{\mathcal{Z}}
\newcommand{\hj}{\hat{j}}
\newcommand{\bAi}{\left(\bA^{-1}\right)}

\section{Introduction and Summary}
It is well known that string theories are not only theories of strings but
also contain number of objects with different dimensionality and properties
\footnote{For recent review, see \cite{Blumenhagen:2013fgp,Becker:2007zj,Polchinski:1998rr}.}.
D-branes are examples of these objects that are in some sense exceptional since
they can be exactly defined "as the planes where open strings can end" and hence
they have exact two dimensional conformal field theory description at least at the
weak coupling regime \cite{Polchinski:1995mt}. Characteristic properties of these
objects (among others) is an existence of the constant known as p-brane tension $T_p$ which
is the mass  per unit of spatial $p-$dimensional volume. The value of the tensions
for different objects can be determined by different methods, see for example
excellent book \cite{Polchinski:1998rr}. On the other hand Dirac-Born-Infeld
(DBI) action for Dp-brane is highly non-linear due to the square root of the determinant.
Further, this action cannot describe objects with zero tension. This problem can
be solved by introducing auxiliary fields so that we replace square-root structure of
the action with the more tractable one when we introduce scalar auxiliary field. Then we can
go even further and replace the constant $T_p$ by $p-$form gauge potential in such
a way that the tension arises as  solution of the equations of motion for this non-dynamical
$p-$form \cite{Bergshoeff:1992gq,Townsend:1997kr,Bergshoeff:1998ha}. This is very attractive
form of the action that is also scale invariant. It is also important to stress that
the integration constant can have  any value so that  it describes
tensionless and negative tension branes
\cite{Dijkgraaf:2016lym} as well, at least in principle. On the other hand the case of the
zero integration constant is again tricky as the tensionless limit of ordinary
Dirac-Born-Infeld action since it implies that the matrix is singular and
hence the derivation of the equation of motion that is valid in case of the non-zero tension
cannot be applied here. The correct procedure how to analyze zero tension solution
is to switch to the Hamiltonian formalism.

Since the idea of the Dp-brane with dynamical tension is very attractive we can try to
extend this construction to the case of an unstable non-BPS Dp-brane
\cite{Sen:1999md,Bergshoeff:2000dq,Garousi:2000tr,Kluson:2000iy}
\footnote{For review, see \cite{Sen:2004nf}.} which is the goal of this paper. We propose
an action for non-BPS Dp-brane with variable tension. Then we determine corresponding equations
of motion. We show that the tachyon kink is the solution of these
equations of motion on condition that the solution of the equation of motion for
$p-$form is non-zero constant. Then we argue,  following \cite{Sen:2003tm,Kluson:2005fj}
that this kink on the world-volume of unstable Dp-brane corresponds to the
$D(p-1)-$brane. It is clear that the equations of motion and kink solution
are not valid in case  when the solution of the equation of motion
 for $p-$form is zero constant.  In order to analyze
this problem we proceed to the Hamiltonian formulation of a non-BPS Dp-brane with
variable tension. We find corresponding Hamiltonian and determine algebra of constraints.
Then we analyze two situations. The first one when the tachyon is sitting at its
minimum value. We argue that the resulting equations of motion with constant electric
flux correspond to the equations of motion derived from the Nambu-Gotto action. In other words
the dynamics of the non-BPS Dp-brane at the tachyon vacuum is equivalent to the
dynamics of fundamental string that is delocalized along the world-volume of
non-BPS Dp-brane
\footnote{For previous works where the fate of the non-BPS Dp-brane at the tachyon
vacuum was analyzed, see  \cite{Sen:2003bc,Kwon:2003qn,Sen:2000kd,Gibbons:2000hf,Lindstrom:2001qa}.}.
We mean that the fact that this solution is delocalized along the world-volume of
an unstable D-brane has a natural interpretation. We know that at the end of the tachyon
vacuum unstable D-brane disappears so that we mean that
it does not make sense to speak about
the position of the string remnant on it. Further, the only physical meaning
has the localization of the string in the target space-time and the dynamics of
these modes is governed  by the equations of motion that follow from Nambu-Gotto action.
We also discuss the second class of the solution of the Hamiltonian equations
of motion for  non-BPS Dp-brane with variable tension when the tension
of this brane is equal to zero. The similar situation was analyzed previously in
\cite{Lindstrom:1999tk,Gustafsson:1998ej,Lindstrom:1997uj,Hassani:1994rf} however with
slightly different limiting procedure. More precisely, the limiting procedure
introduced in \cite{Lindstrom:1999tk,Gustafsson:1998ej,Hassani:1994rf,Lindstrom:1997uj}
the factor in front of $F_{\mu\nu}=\partial_\mu V_\nu-\partial_\nu V_\mu$ scales as well
which is not the case of the non-BPS Dp-brane with the variable tension where the
gauge field has the dimension of length. This follows from the fact that the
tension is generated dynamically as the solution of the equations of motion. On the other
hand there is no way how to generate length scale in front $F_{\mu\nu}$ dynamically.
As a result the zero tension solution of the non-BPS Dp-brane equation of motion
has  similar form as the tachyon vacuum solution with the difference that now
the world-volume theory contains additional massless mode which was the original tachyon
field. In other words the solutions of the equations motion at the zero tension
vacuum correspond to the string propagating in the space with one additional dimension.

All these results are derived in the background with vanishing Ramond-Ramond (RR) fields so that
one can ask the question whether an existence of the background with non-trivial
RR fields does not change the interpretation of the resulting solutions as the solutions
that arise from the Nambu-Gotto action. In other words, if we were found that there is
non-zero coupling to the RR fields at the tachyon vacuum we could not interpret the
resulting configuration as the fundamental string due to the fact that fundamental string
 does not couple to RR fields directly. We demonstrate that this is not the case on the example of
non-BPS D2-brane when we find its Hamiltonian formulation in the presence of the RR fields.
We show that the coupling of this brane to RR fields vanishes at the tachyon vacuum and
hence the resulting configuration really corresponds to the fundamental string. This is
non-trivial result since the Hamiltonian formulation of non-BPS D2-brane in the background
with non-zero RR fields has not been done before.

Let us outline our results. We propose an action for non-BPS Dp-brane with dynamical tension.
We study its properties and show that the equations of motion for this system
has the tachyon kink solution that can be interpreted as a lower dimensional D(p-1)-brane. We also
find the Hamiltonian formulation of p-brane and non-BPS Dp-brane with variable tension
in general background. We study two solutions that cannot be analyzed in the conventional
Lagrangian formalism which are the tachyon vacuum solution and the zero tension solution.
We argue that the solutions of the equations of motion at these vacua and with constant
electric flux have a natural interpretation as the solutions of the Nambu-Gotto equations
of motion which supports the conjecture that at the tachyon vacuum the non-BPS Dp-brane
disappears and the gas of the fundamental string emerges. Finally we also argue that
this conclusion is valid even in the presence of the non-trivial Ramond-Ramond forms.

The organization of this paper is as follows. In the next section (\ref{second}) we
introduce an action for Dp-brane in general background with dynamical tension and we show
that the resulting equations of motion are equivalent to the equations of motion
derived from DBI and WZ D-brane actions. In section (\ref{third}) we perform  Hamiltonian
analysis of p-brane with the variable tension. In section (\ref{fourth})
 we formulate  an action for non-BPS Dp-brane with variable tension and we show that the equations of motion
have a  solution that can be interpreted as D(p-1)-brane. In section (\ref{fifth})
we find Hamiltonian formulation of this Dp-brane and analyze canonical equations of motion.
We also find the Hamiltonian for non-BPS D2-brane with dynamical tension in non-zero
RR  background.

\section{D-brane Action With Variable Tension}\label{second}
In this section we review basic facts about Dp-brane with variable tension
following  \cite{Bergshoeff:1998ha}.
Let us consider the Lagrangian density in the form
\begin{equation}
\mL=\frac{1}{2v}\left[e^{-2\Phi}\det \bA+(\star \mG_{(p+1)})^2\right] \
,
\end{equation}
where $\Phi$ is a dilaton, $v$ is an independent worldvolume
density, and where
\begin{equation}
\bA_{\mu\nu}=g_{\mu\nu}+\mF_{\mu\nu} \ ,
\quad  \mF_{\mu\nu}=\partial_\mu V_\nu-\partial_\nu V_\mu +b_{\mu\nu} \ ,
\end{equation}
where $g_{\mu\nu}$ and $B_{\mu\nu}$ are induced metric and two form respectively
\begin{equation}
g_{\mu\nu}=G_{MN}\partial_\mu X^M\partial_\nu X^N \ , \quad
b_{\mu\nu}=B_{MN}\partial_\mu X^M\partial_\nu X^N \ .
\end{equation}
$G_{MN}$ and $B_{MN}, M, N=0,\dots,9$ are target space-time metric
and NS-NS two form fields respectively, $X^M(\xi)$ are world-volume fields
that parameterize the position of Dp-brane in the target space-time. Finally $V_\mu, \mu=0,\dots,p$
is world-volume gauge field where the world-volume of Dp-brane is labeled with
coordinates $\xi^\mu,\mu=0,\dots,p$.

An important building block of Dp-brane with the variable tension is
the scalar density $\star \mG_{(p+1)}$ which  is
the world-volume Hodge dual of $(p+1)$-form field strength
$\mG_{(p+1)}$ and that has an explicit form
\begin{eqnarray}
\star
\mG_{(p+1)}=\frac{1}{(p+1)!}\epsilon^{\mu_1\dots\mu_{p+1}}\partial_{[\mu_1}
A_{\mu_2\dots\mu_{p+1}]}-\sum_{n\geq 0}
\frac{1}{n! (2!)^n q!}\epsilon^{\mu_1\dots\mu_{p+1}}
(\mF)^n_{\mu_1\dots\mu_{2n}}C_{\mu_{2n+1}\dots\mu_{p+1}} \ ,  \nonumber \\
\end{eqnarray}
where $q=p+1-2n$ and where
\begin{equation}
C_{\mu_1\dots \mu_p}=C_{M_1\dots M_p}\partial_{\mu_1}X^{M_1}\dots
\partial_{\mu_p}X^{M_p}
\end{equation}
is the pull-back of the Ramond-Ramond forms to the world-volume of Dp-brane.

Our goal is to show that the equations of motion that follow from the action
$S=\int d^{p+1}\xi \mL$ are equivalent to the equations of motion that follow from the DBI and WZ actions for ordinary Dp-brane.

To begin with note that the equation of motion for $A$ implies that
\begin{equation}\label{eggen1}
\star \mG_{(p+1)}=Tv \ ,
\end{equation}
where $T$ is a constant. In what follows we will presume that it is not equal to zero.
On the other hand the equation of motion for $v$ implies
\begin{equation}\label{eqgen2}
\star\mG_{(p+1)}=e^{-\Phi} \sqrt{-\det \bA}
\end{equation}
that together with (\ref{eggen1}) implies
\begin{equation}\label{vT}
 v=\frac{1}{T}
e^{-\Phi}\sqrt{-\det \bA}  \ .
\end{equation}
Finally we analyze the equations of motion for $X^M$ and $V_\alpha$.
In case of $X^M$ we obtain
\begin{eqnarray}\label{eqgen3}
& &\frac{1}{2v}\partial_M[e^{-2\Phi}] \det \bA+
\frac{1}{2v}e^{-2\Phi}\left(\partial_M G_{KL}\partial_\alpha X^K
\partial_\beta X^L+\partial_M B_{KL}\partial_\alpha X^K
\partial_\beta X^L\right)(\bA^{-1})^{\beta\alpha}\det \bA\nonumber
\\
&-&\partial_\alpha\left[\frac{1}{v}e^{-2\Phi}G_{MN}\partial_\beta X^N
(\bA^{-1})^{\beta\alpha}_S\det \bA\right]-
\partial_\alpha \left[\frac{1}{v}e^{-2\Phi}B_{MN}\partial_\beta X^N
(\bA^{-1})^{\beta\alpha}_A\det \bA\right]+J_M=0 \ , \nonumber \\
\end{eqnarray}
where
\begin{equation}\label{JM}
J_M=\int d^{p+1}\xi\frac{1}{v}\frac{\delta (\star \mG_{(p+1)})}{\delta X^M}
\star \mG_{(p+1)} \ ,
\end{equation}
and where
\begin{equation}
(\bA^{-1})^{\alpha\beta}_S=\frac{1}{2}
\left((\bA^{-1})^{\alpha\beta}+(\bA^{-1})^{\beta\alpha}\right) \ ,
\quad
(\bA^{-1})^{\alpha\beta}_A=\frac{1}{2}
\left((\bA^{-1})^{\alpha\beta}-(\bA^{-1})^{\beta\alpha}\right) \ .
\end{equation}
Further, the equation of motion for $V_\alpha$ has the form
\begin{eqnarray}
-\partial_\beta\left[\frac{1}{2v}e^{-2\Phi}(\bA^{-1})^{\alpha\beta}_A\det
\bA\right]+J^\alpha=0 \ , \nonumber \\
\end{eqnarray}
where
\begin{equation}\label{Ja}
J^\alpha=\int d^{p+1}\xi \frac{1}{v}\frac{\delta (\star \mG_{(p+1)})}{\delta
V_\alpha}\star\mG_{(p+1)} \ .
\end{equation}
Now we easily see that these equations of motion have the same form
as the equation of motion derived from standard DBI and WZ action.
In fact, inserting (\ref{eggen1}) into (\ref{JM}) and (\ref{Ja})
we obtain
\begin{equation}\label{JWZ}
J^{WZ}_M=T\int d^{p+1}\xi\frac{\delta (\star \mG_{(p+1)})}{\delta X^M} \ , \quad
J^\alpha_{WZ}=T\int d^{p+1}\xi \frac{\delta (\star \mG_{(p+1)})}{\delta
V_\alpha}
\end{equation}
that coincide with the currents introduced in
\cite{Skenderis:2002vf}. Further, inserting
(\ref{eqgen2}) into (\ref{eqgen3}) we obtain the equations of motion
\begin{eqnarray}\label{eqgenDBI}
&T&\partial_M\Phi e^{-\Phi}\sqrt{-\det \bA}-
Te^{-\Phi}\left(\partial_M G_{KL}\partial_\alpha X^K
\partial_\beta X^L+\partial_M B_{KL}\partial_\alpha X^K
\partial_\beta X^L\right)(\bA^{-1})^{\beta\alpha}\sqrt{-\det \bA}\nonumber
\\
&+&\partial_\alpha[Te^{-\Phi}g_{MN}\partial_\beta X^N
(\bA^{-1})^{\beta\alpha}_S\sqrt{-\det \bA}]+
\partial_\beta [Te^{-\Phi}b_{MN}\partial_\beta X^N
(\bA^{-1})^{\beta\alpha}_A\sqrt{-\det \bA}]+J_M^{WZ}=0 \ , \nonumber \\
\end{eqnarray}
which are the equations of motion derived from the DBI and WZ
action. In the same way we proceed with the equation of motion for
$V_\alpha$
\begin{equation}\label{eqVDBI}
T\partial_\beta[e^{-\Phi}(\bA^{-1})^{\alpha\beta}_A\sqrt{-\det\bA}]+J^\alpha_{WZ}=0
\end{equation}
which is the equation of motion derived from DBI action. It is important to stress that given analysis is valid for $T\neq 0$. To see this explicitly note that (\ref{vT}) implies that $\det \bA=0$ for $T=0$ and hence it is not possible to introduce inverse matrix to $\bA$. In order to deal with this case
it is convenient to proceed to the Hamiltonian formulation.
\section{Hamiltonian Formalism for $p-$Brane With Dynamical Tension}\label{third}
In this section we perform the Hamiltonian formulation of p-brane with variable tension. For simplicity we consider the case of pure p-brane leaving the general analysis
to the case of non-BPS Dp-brane that will be performed in the next section.
 In  case of p-brane
 we have
\begin{equation}
\star\mG_{(p+1)}
=\frac{1}{p!}\epsilon^{\mu\mu_2\dots\mu_{p+1}}
\partial_\mu A_{\mu_2\dots\mu_{p+1}}=\partial_\mu \omega^\mu \ ,
\end{equation}
where
\begin{equation}
\omega^\mu=\frac{1}{p!}\epsilon^{\mu\mu_1\dots\mu_p}A_{\mu_1\dots\mu_p}
\
\end{equation}
is vector density of unit weight. With the help of $\omega^\mu$ we can
write the action for p-brane with variable tension in the form
\begin{equation}\label{pHam}
S=\int d^{p+1}\xi \frac{1}{2v}( \det g+(\partial_\mu \omega^\mu)^2) \ .
\end{equation}
Now we proceed to the canonical formulation of this theory.
From (\ref{pHam}) we derive conjugate momenta
\begin{eqnarray}\label{pmp}
p_M&=&\frac{\delta L}{\delta \partial_0 X^M}=
\frac{1}{v}G_{MN}\partial_\mu X^N g^{\mu 0}\det g \ , \nonumber \\
p_v&\approx & 0 \ , \quad  \tau_0=\frac{\delta L}{\delta \partial_0 \omega^0}=
\frac{1}{v}\partial_\mu \omega^\mu \ , \quad \tau_i=\frac{\delta L}{\delta
\partial_0 \omega^i}\approx 0 \ . \nonumber \\
\end{eqnarray}
Then the bare  Hamiltonian density is equal to
\begin{eqnarray}
\mH_B&=&p_M\partial_0 X^M+\rho_0\partial_0 \omega^0-\mL
\nonumber \\
&=&\frac{v}{2\det g_{ij}}p_M G^{MN}p_N+\frac{1}{2}\tau_0^2v-\partial_i\omega^i\tau_0 \nonumber \\
\end{eqnarray}
using the fact that
\begin{equation}
p_M G^{MN}p_N=
\frac{1}{v^2}\det g_{ij}\det g \ .
\end{equation}
Further, from (\ref{pmp}) we derive following
 primary constraints
\begin{equation}
\mH_i=p_M\partial_i X^M \approx 0 \ .
\end{equation}
As the result the  Hamiltonian density  with all primary constraints included has the
form
\begin{equation}
\mH_E=\frac{v}{2\det g_{ij}}p_M G^{MN}p_N+\frac{1}{2}\tau_0^2v-\partial_i\omega^i\tau_0+N^i\mH_i+U_v
p_v+U^i\rho_i \ .
\end{equation}
Now we have to study the stability of the primary constraints when the Hamiltonian that generates the time evolution is $H_E=\int d^p\xi \mH_E$. The
requirement of the preservation of the constraint $p_v\approx 0$
implies
\begin{eqnarray}
\partial_t p_v=\pb{p_v,H_E}=
-\frac{1}{2\det g_{ij}}\mH_0 \ , \quad \mH_0=p_M
G^{MN}p_N+\tau_0^2\det g_{ij} \
 \nonumber \\
 \end{eqnarray}
 while the preservation of the constraint $\rho_i\approx 0$ implies an existence of additional constraints
 \begin{equation}
\partial_0 \rho_i=\pb{\rho_i,H_E}=\partial_i \tau^0\equiv \mG_i\approx 0 \ .
\end{equation}
Finally we proceed to the analysis of the preservation of the
constraints $\mH_i$. We extend these constraints with  the secondary constraints
$\mG_i$ in order to ensure that they are
 preserved during the time evolution.
In more details, let us introduce the constraints
\begin{equation}
\tmH_i=\mH_i-\mG_i\omega^0
\end{equation}
and its smeared form
\begin{equation}
\bH_S(N^i)=\int d^p\xi N^i\tmH_i \ .
\end{equation}
Now it is easy to see that this constraint
has following non-zero Poisson brackets
\begin{equation}
\pb{\bH_S(N^i),p_M}=-\partial_i (N^i p_M) \ , \quad \pb{\bH_S(N^i),X^M}
=-N^i\partial_i X^M \ , \quad  \pb{\bH_S(N^i),\tau_0}=-N^i\partial_i\tau_0 \
\end{equation}
and hence we find
\begin{eqnarray}
\pb{\bH_S(N^i),\mH_0}=-2\partial_i N^i\mH_0-N^i\partial_i \mH_0  \ ,
\pb{\bH_S(N^i),\bH_S(M^j)}=\bH_S(N^j\partial_j M^i-\partial_j N^i M^j)
\nonumber \\
\end{eqnarray}
which implies that $\tmH_i$ are preserved during the time evolution of the system.
Now in order to find the total Hamiltonian we have to include all
constraints to it. We absorb the factor
$\frac{v}{\det g_{ij}}$ into the Lagrange multiplier $N_0$
corresponding to the constraint $\mH_0$. In the same way we
include $\omega^i$ into the definition of the Lagrange multiplier $\Gamma^i$
corresponding to the constraint $\mG_i$. As a result the total Hamiltonian has the
form
\begin{equation}
H_T=\int d^p\xi (N_0 \mH_0+N^i\mH_i+\Gamma^i\mG_i) \ ,
\end{equation}
where we do not induced the constraints $p_v\approx 0 \ ,
\rho^i\approx 0$ since they decouple from the theory.  Let us now
consider the equation of motion for $
\tau_0$
\begin{equation}
\partial_0 \tau_0=\pb{\tau_0,H_T}=0 \
\end{equation}
since $H_T$ does not depend on $\omega^0$.
Further, the constraint $\mG_i\approx 0$ implies that
$\partial_i\tau_0=0$ and consequently we find that $\tau_0=T$ is a
constant that can be identified with  the tension of p-brane. Let us now determine
remaining equations of motion
\begin{eqnarray}\label{eqpHam}
\partial_0 X^M&=&\pb{X^M,H_T}=2N_0G^{MN}p_N+N^i\partial_i X^M \ ,
\nonumber \\
\partial_0 p_M&=&\pb{p_M,H_T}=-N_0\partial_M G^{KL}p_Kp_L
\nonumber \\
&-&N_0\tau_0^2 \partial_M
G_{KL}\partial_i X^K\partial_j X^L g^{ji}\det g_{ij}
+2\partial_i \left[N_0\tau_0^2 G_{MN}\partial_j X^N g^{ji}\det g_{ij}\right]
-\partial_i (N^ip_M) \ .
\nonumber \\
\end{eqnarray}
We see that for $\tau_0=0$ the equations of motion
(\ref{eqpHam}) simplify considerably
\begin{eqnarray}
\partial_0 X^M&=&\pb{X^M,H_T}=2N_0G^{MN}p_N+N^i\partial_i X^M \ ,
\nonumber \\
\partial_0 p_M&=&\pb{p_M,H_T}=-N_0\partial_M G^{KL}p_Kp_L
-\partial_i (N^ip_M) \
\nonumber \\
\end{eqnarray}
together with the constraints $p_M\partial_i X^M\approx 0 \ ,
p_M G^{MN}p_N\approx 0$. Clearly the diffeomorphism constraint
 can be solved by imposing $X^M=X^M(\xi^0)$. On the other hand
we can consider more general dependence of
\begin{equation}
p_M=P_N(\xi^0)f(\xi^1,\dots,\xi^p) \ , \quad
N_0=n(\xi^0)f^{-1}(\xi^1,\dots,\xi^p) \ , \quad N^i=n^i(\xi^0)
f^{-1}(\xi^1,\dots,\xi^p)
\end{equation}
where $f(\xi^1,\dots,\xi^p)$ is arbitrary function
\cite{Sen:2000kd}. Then it is easy to see that
 the equation of motion for $X^M$ and $P_M$
 correspond to the equation of motion for
 tensionless particle where the localization of the particle along
 the world-volume of p-brane is determined by the function
 $f(\xi^1,\dots,\xi^p)$. On the other hand the physical
 meaning has the localization of this object in the target
 space time where the dynamics of the embedding modes is
 governed by the equations of motions for  massless particle.
\section{Non-BPS Dp-brane with Variable Tension}\label{fourth}
In this section we propose an action for  non-BPS Dp-brane with variable
tension. We claim that the Lagrangian density  has the form
\begin{equation}
\mL_{non}=\frac{1}{2v}\left[e^{-2\Phi}V^2(T)\det \tbA+(\star
G_{(p+1)})^2\right] \ ,
\end{equation}
where $\Phi$ is a  dilaton, $v$ is an independent worldvolume
density  and where
\begin{equation}
\tbA_{\mu\nu}=G_{MN}\partial_\mu X^M\partial_\nu X^N+\partial_\mu
T\partial_\nu T+ \mF_{\mu\nu} \ ,
\end{equation}
where $\mF_{\mu\nu}=B_{MN}\partial_\mu X^M\partial_\nu X^N+
(\partial_\mu V_\nu-\partial_\nu V_\mu)$. Further, $T$ is
the tachyon field and $V(T)$ is a corresponding potential that is
symmetric under $T\rightarrow -T$ and it has a maximum at $T=0$ and
has its minimum at $T=\pm \infty$ where it vanishes
\cite{Sen:2004nf}.

In case of the non-BPS Dp-brane we propose
the scalar density  $\star{\mG}$  in the form
\begin{eqnarray}
\star
\tmG=\frac{1}{(p+1)!}\epsilon^{\mu_1\dots\mu_{p+1}}\partial_{[\mu_1}
A_{\mu_2\dots\mu_{p+1}]}-\sum_{n\geq 0} \frac{1}{n! (2!)^n
q!}\epsilon^{\mu_1\dots\mu_{p+1}}V(T)
(\mF)^n_{\mu_1\dots\mu_{2n}}C_{\mu_{2n+1}\dots\mu_{p}}\partial_{\mu_{p+1}}T \nonumber \\
\end{eqnarray}
where $q=p+1-2n-1$. Now we show that the equations of motion derived
from the action $S_{non}=\int d^{p+1}\xi \mL_{non}$ have the same form as the equations of motion
derived from non-BPS Dp-brane action. First of all the equation of
motion with respect to $A$ implies
\begin{equation}\label{eq1T}
\frac{\star \tmG}{v}=\tau_p \ ,
\end{equation}
where $\tau_p$ can be interpreted as non-BPS Dp-brane tension.
On the other hand
 the equation of motion with respect to $v$ implies
\begin{equation}\label{eq2T}
e^{-2\Phi}V^2 \det \tbA+(\star \tmG)^2=0  \
\end{equation}
that with the help of (\ref{eq1T}) implies (on condition that
$\tau_p\neq 0$)
\begin{equation}\label{vsolT}
v=\frac{1}{\tau_p}e^{-\Phi}V(T) \sqrt{-\det \bA} \ .
\end{equation}
Finally we analyze the equations of motion for $X^M$ and $V_\alpha$
and $T$.  For the first one we obtain
\begin{eqnarray}\label{eq3T}
& &\frac{1}{2v}\partial_M[e^{-2\Phi}] V^2 \det \tbA+
\frac{1}{2v}e^{-2\Phi}V^2\left(\partial_M G_{KL}\partial_\alpha X^K
\partial_\beta X^L+\partial_M B_{KL}\partial_\alpha X^K
\partial_\beta X^L\right)(\tbA^{-1})^{\beta\alpha}\det \tbA\nonumber
\\
&-&\partial_\alpha[\frac{1}{v}e^{-2\Phi}V^2G_{MN}\partial_\beta X^N
(\tbA^{-1})^{\beta\alpha}_S\det \tbA]-
\partial_\beta [\frac{1}{v}V^2e^{-2\Phi}B_{MN}\partial_\beta X^N
(\tbA^{-1})^{\beta\alpha}_A\det \tbA]+\tJ_M=0 \ , \nonumber \\
\end{eqnarray}
where
\begin{eqnarray}\label{JMnon}
\tJ_M&=&\int d^{p+1}\xi\frac{1}{v}\frac{\delta (\star \tmG)}{\delta
X^M} \star \tmG= \nonumber \\
&- &\sum_{n\geq 0}\frac{1}{n! (2!)^n q!} \epsilon^{\mu_1\dots\mu_{p+1}}
\left(\frac{\star\tmG}{v}V(T)n \partial_M
B_{KL}\partial_{\mu_1}X^K\partial_{\mu_2}X^L
(\mF)^{n-1}_{\mu_3\dots\mu_{2n}}C_{\mu_{2n+1}\dots\mu_p}\partial_{\mu_{p+1}}T
\right.\nonumber \\
&+&\frac{\star\tmG}{v}V(T) (\mF)^n_{\mu_1\dots\mu_{2n}}\partial_M
C_{K_1\dots K_q}\partial_{\mu_{2n+1}}X^{K_1}\dots
\partial_{\mu_q}X^{K_q}\partial_{\mu_{p+1}}T
\nonumber \\
&-&2n\partial_{\mu_1}[\frac{\star\tmG}{v}V(T)b_{MK}\partial_{\mu_2}X^K
(\mF)^{n-1}_{\mu_3\dots
\mu_{2n}}C_{\mu_{2n+1}\dots\mu_p}\partial_{\mu_{p+1}}T]\nonumber \\
&-& \left. q\partial_{2n+1}[\frac{\star\tmG}{v}V(T)(\mF)^n_{\mu_1\dots\mu_{2n}}C_{MK_2\dots
K_q}
\partial_{\mu_{2n+2}}X^{M_2}\dots\partial_{\mu_p}X^{K_q}\partial_{\mu_{p+1}}T]\right) \ .
\nonumber \\
\end{eqnarray}
Further, the equations of motion for $V_\alpha$ have the form
\begin{eqnarray}\label{eqTV}
-\partial_\beta\left[\frac{1}{v}e^{-2\Phi}V^2(\tbA^{-1})^{\alpha\beta}_A\det
\tbA\right]+\tJ^\alpha=0 \ , \nonumber \\
\end{eqnarray}
where
\begin{eqnarray}\label{Janon}
\tJ^\alpha&=&\int d^{p+1}\xi \frac{1}{v}\frac{\delta (\star
\tmG)}{\delta V_\alpha}\star\tmG\nonumber \\
&=&-\sum_{n\geq 0}\frac{2n}{n! (2!)^n
q!}\partial_{\mu_2}[\frac{\star\mG}{v}\epsilon^{\alpha\mu_2\dots\mu_{p+1}}(\mF)^{n-1}_{\mu_3\dots\mu_{2n}}
C_{\mu_{2n+1}\dots\mu_p}\partial_{\mu_{p+1}}T] \ .
\end{eqnarray}
Finally the equation of motion for $T$ has the form
\begin{eqnarray}\label{eq4T}
e^{-2\Phi}\frac{dV}{dT}V\det
\tbA-\partial_\alpha\left[\frac{1}{v}e^{-2\Phi}V^2\partial_\beta T
(\tbA^{-1})^{\beta\alpha}_S\det \tbA\right]+\tJ_T=0 \ , \nonumber \\
\end{eqnarray}
where
\begin{eqnarray}\label{JTnon}
\tJ_T&=&\int d^{p+1}\xi \frac{1}{v}\frac{\delta (\star \tmG)}{\delta
T} \star \tmG=\nonumber \\
&=&\sum_{n\geq 0}\frac{1}{n! (2!)^n
q!}\epsilon^{\mu_1\dots\mu_{p+1}}\left(\frac{\star\tmG}{v}V'(T) (\mF)^n_{\mu_1\dots\mu_{2n}}
C_{\mu_{2n+1}\dots \mu_p}\partial_{\mu_{p+1}}T\right.\nonumber \\
&-&\left.\partial_{\mu_{p+1}}[\frac{\star\mG}{v}V(T)
(\mF)^n_{\mu_1\dots\mu_{2n}}
C_{\mu_{2n+1}\dots \mu_p}]\right) \nonumber \\
 \end{eqnarray}
Now we show that the  equations of motion (\ref{eq1T}),(\ref{eq2T}),
(\ref{eq3T}) and (\ref{eq4T}) can be solved with tachyon kink ansatz
that can be interpreted as a lower dimensional D(p-1)-brane, following
the similar analysis presented in case of non-BPS Dp-brane
in \cite{Sen:2003tm,Kluson:2005fj}.
 We
choose one world-volume coordinate, say $\xi^p\equiv x$ and consider
following ansatz for the tachyon
\begin{equation}\label{Tanst}
T(x,\xi^{\hmu})=f(a(x-t(\xi^{\hmu})) \ ,
\end{equation}
where as in \cite{Sen:2003tm} we presume that $f(u)$ satisfies following
properties
\begin{equation}
f(-u)=-f(u) \ , \quad f'(u)>0 \ , \forall u \ , \quad f(\pm
\infty)=\pm \infty \ ,
\end{equation}
which is however an arbitrary function of its argument $u$. $a$ is a
constant which can be taken to $\infty$ in the end that leads to the
configuration when $T=\infty$ for $x>t(\xi^{\hmu})$ and $T=-\infty$
for $x<t(\xi^{\hmu})$. Finally note that $\hmu=0,\dots,p-1$. Let us
also presume following ansatz for massless fields
\begin{equation}\label{Xanst}
X^M(x,\xi^{\hmu})=X^M(\xi^{\hmu}) \ , \quad  A_x(x,\xi^{\hmu})=0 \ , \quad
A_{\hmu}(x,\xi^{\hmu})=A_{\hmu}(\xi^{\hmu}) \ .
\end{equation}
Our goal is to show that the dynamics of the kink is governed by D(p-1)-brane  equations
of motion derived in  section (\ref{second}). As the first step  we determine the matrix $\tbA$ for the ansatz
(\ref{Tanst}) and (\ref{Xanst})
\begin{equation}
\tbA_{\mu\nu}=\left(\begin{array}{cc} \bA_{\hmu \hnu}+a^2 f'^2
\partial_{\hmu}t\partial_{\hnu}t & -a^2f'^2\partial_{\hnu}t \\
-a^2 f'^2 \partial_{\hmu} t & a^2 f'^2 \\ \end{array}\right)
\end{equation}
where
\begin{equation}\label{bAans}
\bA_{\hmu\hnu}=G_{MN}\partial_{\hmu}X^M \partial_{\hnu}X^N+
\mF_{\hmu\hnu} \ , \quad  \mF_{\hmu\hnu}=B_{MN}\partial_{\hmu}X^M
\partial_{\hnu}X^N+
(\partial_{\hmu}V_{\hnu}-\partial_{\hnu}V_{\hmu}) \ .
\end{equation}
Note that for the matrix (\ref{bAans}) the determinant $\det\tbA$ has
a form
\begin{equation}
\det \tbA=\det (\tbA_{\hmu\hnu}-\tbA_{\hmu x}\frac{1}{\tbA_{xx}}
\tbA_{x\hnu})\tbA_{xx}=a^2 f'^2\det \bA_{\hmu\hnu} \ .
\end{equation}
As the next step we  determine inverse  matrix
$\tbA^{-1}$
that has following exact form (for all $a$)
\begin{eqnarray}\label{matinverse}
(\tbA^{-1})^{\hmu\hnu}&=&(\bA^{-1})^{\hmu\hnu} \ , \quad
 (\tbA^{-1})^{xx}=\frac{1+a^2f'^
2\partial_{\hmu}t(\bA^{-1})^{\hmu\hnu}\partial_{\hnu}t}{a^2f'^2}  \ , \nonumber \\
(\tbA^{-1})^{\hmu x}&=&(\bA^{-1})^{\hmu \hnu}\partial_{\hnu}t \ ,
\quad (\tbA^{-1})^{x\hnu }=\partial_{\hmu}t(\bA^{-1})^{\hmu \hnu} \
.
\nonumber \\
\end{eqnarray}
Now we show  that the ansatz (\ref{Tanst}) and
(\ref{Xanst}) is solution of the equation of motion. First of all using
(\ref{matinverse}) we easily find
\begin{eqnarray}
\partial_\mu T (\tbA^{-1})^{\mu \hnu}_S=0 \ , \quad
\partial_\mu T(\tbA^{-1})^{\mu x}_S=
\frac{1}{af'} \ . \nonumber \\
\end{eqnarray}
Then it is easy to see that
the equation of motion for the tachyon is satisfied since
\begin{eqnarray}
\frac{1}{v}e^{-2\Phi}\frac{dV}{dT}V\det
\tbA-\partial_\alpha\left[\frac{1}{v}e^{-2\Phi}V^2\partial_\beta T
(\tbA^{-1})^{\beta\alpha}_S\det \tbA\right]+\tJ_T=0 \ ,  \nonumber \\
\end{eqnarray}
where we also used (\ref{vsolT})
together with the fact that  the current $\tJ_T$ vanishes identically for
the ansatz (\ref{Tanst}) and (\ref{Xanst}).
Now we can proceed to the equations of motion for $X^M$. Using the
fact that $X^M$ depend on $\xi^{\hmu}$ only and the explicit form of the
inverse matrix $\tbA^{-1}$ given in (\ref{matinverse})
 we obtain that the
equation (\ref{eq3T}) has the form
\begin{eqnarray}\label{eq3Ta}
& -&\tau_p Vaf'\partial_M[e^{-\Phi}] \sqrt{-\det \bA_{\hmu\hnu}}\nonumber \\
&-&af'\tau_pV \frac{1}{2}e^{-\Phi}\left(\partial_M
G_{KL}\partial_{\hmu} X^K
\partial_{\hnu} X^L+\partial_M B_{KL}\partial_{\hmu} X^K
\partial_{\hnu} X^L\right)(\bA^{-1})^{\hnu\hmu}\sqrt{-\det \bA_{\hmu\hnu}}\nonumber
\\
&+& af'V \tau_p\partial_{\hnu}[e^{-\Phi}G_{MN}\partial_{\hmu} X^N
(\tbA^{-1})^{\hmu\hnu}_S\sqrt{-\det \bA_{\hmu\hnu}}]+
af'V\partial_{\hnu} [e^{-\Phi}B_{MN}\partial_{\hmu} X^N
(\tbA^{-1})^{\hmu\hnu}_A\sqrt{-\det \bA_{\hmu\hnu}}]
\nonumber \\
&+&\tJ_M=0 \ , \nonumber \\
\end{eqnarray}
where we also used the fact that
for any function $F(a(x-t(\xi^{\hmu})))$ we have
\begin{equation}
\partial_{\hmu}F(a(x-t(\xi^{\hmu}))=-\partial_x F(a(x-t(\xi^{\hmu})))\partial_{\hmu}t \ .
\end{equation}
Let us now calculate $\tJ_M$. It is easy to see that the only non-zero contribution
contains $x-$derivative of $T$ since in the opposite case the presence of the
totally antisymmetric tensor $\epsilon$ implies $x-$derivative of massless fields
which are zero by definition. Then $\tJ^M$ is equal to
\begin{equation}
\tJ^M=Vaf'J^M_{WZ} \ ,
\end{equation}
where $J^M_{WZ}$ is the current for D(p-1)-brane (\ref{JWZ}). Collecting
all these results together we obtain that the equations
of motion for $X^M$ have the form
\begin{eqnarray}\label{eqMfinalans}
& &Vaf'\left(-\tau_p \partial_M[e^{-\Phi}]
\sqrt{-\det \bA_{\hmu\hnu}}\right.\nonumber \\
&-&\tau_p \frac{1}{2}e^{-\Phi}\left(\partial_M
G_{KL}\partial_{\hmu} X^K
\partial_{\hnu} X^L+\partial_M B_{KL}\partial_{\hmu} X^K
\partial_{\hnu} X^L\right)(\bA^{-1})^{\hnu\hmu}\sqrt{-\det \bA_{\hmu\hnu}}\nonumber
\\
&+&\ \left. \tau_p\partial_{\hnu}[e^{-\Phi}G_{MN}\partial_{\hmu} X^N
(\tbA^{-1})^{\hmu\hnu}_S\sqrt{-\det \bA_{\hmu\hnu}}]+
\partial_{\hnu} [e^{-\Phi}B_{MN}\partial_{\hmu} X^N
(\tbA^{-1})^{\hmu\hnu}_A\sqrt{-\det \bA_{\hmu\hnu}}]+J^{WZ}_M\right)=0 \  \nonumber \\
\end{eqnarray}
with following physical interpretation. Since for large $T$ the potential $V(T)$
goes as $V\sim e^{-T^2}$ we easily see that
 $aV$ is zero for all points $x\neq t(\xi^{\hmu})$ in the limit $a\rightarrow \infty$.
In other words the kink is localized at the point $x=t(\xi^{\hmu})$. It is important to stress
that $t(\xi^{\hmu})$ does not have any physical meaning since we consider manifestly world-sheet diffeomorphism
invariant theory and hence this kink can be localized at any point on the world-volume of non-BPS
Dp-brane. On the other hand we see that at the point $x=t(\xi^{\hmu})$ the equation of motion for $X^M$
are satisfied on condition when the expression in the bracket $(\dots)$ is zero. However this expression
 is exactly the
equations of motion (\ref{eqgenDBI}) for D(p-1)-brane.

In the same way we proceed with the equation of motion for $V_x$ and $V_{\hmu}$. In the first
case we find that the equation of motion (\ref{eqTV}) has the form
\begin{equation}\label{eqTVa1}
af'V\partial_{\hnu}t\left(
\partial_{\hmu}[e^{-\Phi}(\bA^{-1})^{\hmu\hnu}\sqrt{-\det
\bA_{\hmu\hnu}}]+J^{\hnu}_{WZ}\right)=0 \ ,
\end{equation}
where we used the fact that $\tJ^x$ evaluated on the ansatz (\ref{Tanst}),(\ref{Xanst})
is equal to
\begin{eqnarray}
\tJ^x=
-af'V\sum_{n\geq 0}
\frac{2n}{(n)!(2!)^{n}q!}\epsilon^{\hnu\hmu_2\dots\hmu_{p}x}
\partial_{\hmu_2}[(\mF)^{n-1}_{\hmu_3\dots\hmu_{2n}}
C_{\hmu_{2n+1}\dots\hmu_p}]\partial_{\hnu}t
=af'VJ^{\hnu}_{WZ}\partial_{\hnu}t \ .  \nonumber \\
\end{eqnarray}
In the similar way we can proceed with $V_{\hmu}$ and
we find that the equation (\ref{eqTV}) has the form
\begin{eqnarray}\label{eqTVa2}
\tau_p
Vaf'\left(\partial_{\hnu}[e^{-\Phi}(\bA^{-1})^{\hmu\hnu}_A\sqrt{-\det
\bA_{\hmu\hnu}}]+J^{\hmu}_{WZ}\right)=0 \ .  \nonumber \\
\end{eqnarray}
The physical interpretation of the equations (\ref{eqTVa1}) and
(\ref{eqTVa2}) is the same as  in case of the equation of motion for $X^M$. Explicitly, these equations are
valid at all points $x\neq t(\xi^{\hmu})$ while at $x=t(\xi^{\hmu})$ they
are obeyed on condition that the expression in the bracket is zero. However
this is exactly the equation of motion  (\ref{eqVDBI}) for the gauge field on
the world-volume of D(p-1)-brane. In summary we have found that the tachyon
kink on the world-volume of non-BPS Dp-brane with dynamical tension corresponds
to the lower dimensional stable D(p-1)-brane. It is important to stress that
given analysis is valid on condition that $\tau_p\neq 0$. In order to analyze
solution with $\tau_p=0$ we have to switch to the Hamiltonian formulation
of non-BPS Dp-brane with dynamical tension.
\section{Hamiltonian Formalism for non-BPS Dp-brane with Dynamical
Tension}\label{fifth}
 In order to find solution with $\tau_p=0$ it is
useful to pass to the Hamiltonian formalism. For simplicity we
 presume zero RR background so that the action has the form
\begin{equation}\label{SnonHam}
S=\int d^{p+1}\xi \frac{1}{2v}(e^{-2\Phi}V^2\det \tbA+(\partial_\mu
\omega^\mu)^2) \ .
\end{equation}
From this action we find conjugate momenta
\begin{eqnarray}\label{pMnon}
p_M&=&\frac{1}{v}V^2 e^{-2\Phi}(G_{MN}\partial_\mu X^M(\tbA^{-1})^{\mu
0}_S+B_{MN}\partial_\mu
X^N(\tbA^{-1})^{\mu 0}_A)\det \tbA \ , \nonumber \\
\pi^i&=&\frac{1}{v}V^2 e^{-2\Phi}(\tbA^{-1})^{i0}_A \det \tbA
 \ ,  \quad \pi^0\approx 0 \ ,
 \nonumber \\
 \tau_0&=&\frac{\delta L}{\delta \partial_0
 \omega^0}=\frac{1}{v}\partial_\mu\omega^\mu \ , \quad
 \tau_i=\frac{\delta L}{\delta \partial_0 \omega^i}\approx 0 \ .
 \nonumber \\
 p_T&=&\frac{\delta L}{\delta \partial_0
 T}=\frac{1}{v}e^{-2\Phi}V^2\partial_\beta T (\tbA^{-1})^{\beta 0}_S
 \det \tbA \ . \nonumber \\
 \end{eqnarray}
Using these relations we easily find that the
 bare Hamiltonian is equal to
\begin{eqnarray}
H_B
=\int d^p\xi \left(\frac{e^{-2\Phi}}{2v}V^2 \det
\tbA+\frac{\tau_0^2}{2}v-
\partial_i\omega^i\tau_0+\partial_i V_0\pi^i\right) \ .  \nonumber \\
\end{eqnarray}
In order to express $\det\tbA$ as a function of the canonical
variables we use the fact that
\begin{eqnarray}\label{PiMcondef}
\Pi_M G^{MN}\Pi_N+p_T^2+\pi^i
(\tbA_S)_{ij}\pi^j=
 \frac{e^{-4\Phi}V^2}{v^2} \det
\tbA_{ij} \det \tbA \ ,
 \nonumber \\
\end{eqnarray}
where
\begin{equation}
\Pi_M\equiv p_M-B_{MN}\partial_i X^N \pi^i \ .
\end{equation}
Note also  that  (\ref{pMnon})
imply following primary constraint
\begin{eqnarray}
\mH_i\equiv p_M\partial_i X^M+F_{ij}\pi^j+p_T\partial_i T\approx 0 \ .
\nonumber \\
\end{eqnarray}
Now we are ready to write an extended form of the
Hamiltonian with all primary constraints included
\begin{eqnarray}
H_E=
\int d^p\xi (\frac{ve^{2\Phi}}{2\det \tbA_{ij}V^2}\tmH_0+\omega^i
\partial_i\tau_0+N^i\mH_i-V_0\partial_i\pi^i) \ ,  \nonumber \\
\end{eqnarray}
where
\begin{eqnarray}
\tmH_0=\Pi_M G^{MN}\Pi_N+p_T^2+\pi^i
(\tbA_S)_{ij}\pi^j+V^2\tau_0^2e^{-2\Phi}\det \tbA_{ij} \ . \nonumber \\
\end{eqnarray}
Now the requirement of the preservation of the primary constraints
$p_v\approx 0 \ , \rho_i\approx 0$ and $\pi^0$ implies following
secondary constraints
\begin{eqnarray}
\mH_0\approx 0 \ , \mG_i=\partial_i\tau_0\approx 0 \ , \mG_V=
\partial_i\pi^i\approx 0
\  \nonumber \\
\end{eqnarray}
so that the total Hamiltonian with all constraints included has the form
\begin{equation}\label{HTotalNOn}
H_T=\int d^p\xi (N_0\mH_0+N^i\tmH_i+V^i\mG_i+U^v \mG_V+\rho^v p_v+\rho^i\tau_i) \ ,
\end{equation}
where the extended diffeomorphism constraint has the form
\begin{equation}
\tmH_i=p_N\partial_i X^M+F_{ij}\pi^j +p_T\partial_i T-\partial_i\tau_0\omega^0 \ .
\end{equation}
Finally we have to show that all constraints are preserved during the time
evolution of the system. To do this we introduce  smeared form of these constraints
\begin{eqnarray}
\bH_T(N)&=&\int d^p\xi N \tmH_0 \ , \quad \bH_S(N^i)=
\int d^p\xi N^i\tmH_i \ , \nonumber \\
 \bG_A(V)&=&\int d^p\xi V\mG_V \ , \quad
\bG(W^i)=\int d^p\xi W^i\mG_i  \ . \nonumber \\
\end{eqnarray}
Now with the help of the  following Poisson brackets
\begin{eqnarray}
\pb{\bH_S(N^i),X^M}&=&-N^i\partial_i X^M \ , \quad
\pb{\bH_S(N^i),p_M}=-\partial_i(N^ip_M) \ , \nonumber \\
\pb{\bH_S(N^i),A_i}&=&-N^k F_{ki}  \ , \quad
\pb{\bH_S(N^i),\pi^i}=-\partial_k (N^k\pi^i)+\partial_lN^i\pi^l+N^i\mG_V \ ,
\nonumber \\
\pb{\bH_S(N^i),T}&=&-N^i\partial_i T \ , \quad \pb{\bH_S(N^i),p_T}=-
\partial_i (N^i p_T) \  \nonumber \\
\nonumber \\
\end{eqnarray}
we obtain
\begin{eqnarray}
\pb{\bH_S(N^i),\bH_T(M)}=\bH_T(N^i\partial_i M)+2\bG_A(MN^i(\tbA_S)_{ij}\pi^j+
M\Pi^M B_{MN}\partial_i X^N) \ , \nonumber \\
\end{eqnarray}
and also
\begin{equation}
\pb{\bH_S(N^i),\bH_S(M^j)}=\bH_S(N^j\partial_j M^i-\partial_j N^iM^j)+
\bG_A(M^iF_{ij}N^j) \ .
\end{equation}
Finally we calculate the Poisson bracket $\pb{\bH_T(N),\bH_T(M)}$. Using the fact that
\begin{equation}
\pb{\Pi_M(\xi),\Pi_N(\xi')}=H_{MNK}\partial_i X^K\delta(\xi-\xi')+
B_{MN}\mG \delta(\xi-\xi')
\end{equation}
and after some tedious calculations we obtain the result
\begin{eqnarray}
\pb{\bH_T(N),\bH_T(M)}&=&\bH_S ((N\partial_j M-M\partial_j N)
 \hat{\tbA}^{ji}_S V^2\tau_0^2 e^{-2\Phi}\det\tbA_{ij})+\nonumber \\
&+&\bG( (N\partial_j M-M\partial_j N) \omega^0 \hat{\tbA}_S^{ji}
V^2\tau_0^2 e^{-2\Phi}\det \tbA_{ij})+
\nonumber \\
&+&4\bH_S((N\partial_j M-M\partial_j N)\pi^j \pi^i)
+4\bG_A ((N\partial_j M-M\partial_j N)\pi^i
\pi^j\omega^0) \ .  \nonumber \\
\end{eqnarray}
From the previous Poisson brackets we see that
$\mH_0,\tmH_i,\mG,\mG_i$ are the first class constraints and
no new constraints are generated during the time evolution of the system.
Now we proceed to the solution of the equations of motion for $\tau_0,A_i$ and $\pi^i$ that follow
from the Hamiltonian (\ref{HTotalNOn})
\begin{eqnarray}\label{eqhamtau}
\partial_0\tau_0&=&\pb{\tau_0,H_T}=0 \ ,
\nonumber \\
\partial_0 A_i&=&\pb{A_i,H_T}=-2N_0B_{MN}\partial_i X^N G^{MK}\Pi_K+
(\tbA_S)_{ij}\pi^j+\partial_i V_0+N^jF_{ji}  \ , \nonumber \\
\partial_0\pi^i&=&\pb{\pi^i,H_T}=-2\partial_k[V^2\tau_0^2
e^{-2\Phi}\hat{\tbA}^{ki}_A\det
\tbA_{ij}]+\partial_k[N^k\pi^i-N^i\pi^k] \ , \nonumber \\
\end{eqnarray}
where $\hat{\tbA}^{ki}$ is the matrix inverse to
$\tbA_{ij}\hat{\tbA}^{jk}=\delta_i^k$. Further, the
equations of motion for $T$ and $p_T$ have the form
\begin{eqnarray}\label{eqTham}
\partial_0 T&=&\pb{T,H_T}=2N_0 p_T+N^i\partial_i T \ , \nonumber \\
\partial_0p_T&=&\pb{p_T,H_T}\nonumber \\
&=&2\partial_i(\pi^i\partial_j
T\pi^j)-2V\frac{dV}{dT}\tau_0^2e^{-2\Phi}\det \tbA_{ij}
+\partial_i[V^2 e^{-2\Phi}\partial_j T
\hat{\tbA}^{ji}\det\tbA]+\partial_i [N^i p_T] \ .
\nonumber \\
\end{eqnarray}
Finally we determine  equations of motion for $X^M$ and $p_M$
\begin{eqnarray}
\partial_0 X^M&=&\pb{X^M,H_T}=2N_0 G^{MN}\Pi_N+N^i\partial_i X^M  \ ,
\nonumber \\
\partial_0 p_M&=&\pb{p_M,H_T}=N_0\Pi_K \partial_M G^{KL}
\Pi_L+2N_0\partial_M B_{KL}\partial_i X^L \pi^i G^{KN}\Pi_N-2
\partial_i[N_0B_{KM}\pi^i G^{KN}\Pi_N]\nonumber \\
&-&N_0\pi^i\partial_iX^K\partial_M G_{KL}\partial_jX^L\pi^j+
2\partial_i[N_0\pi^i G_{MN}\partial_j X^N\pi^j]-N_0V^2 \tau_0^2
\partial_M[e^{-2\Phi}]\det \tbA_{ij}-\nonumber \\
&-&N_0V^2e^{-2\Phi}\tau_0^2(\partial_i X^K \partial_M G_{KL}\partial_j
X^L+\partial_i X^K\partial_M B_{KL}\partial_j
X^L)\hat{\tbA}^{ji}\det\tbA_{ij} \nonumber \\
&+&2\partial_i [N_0 V^2 e^{-2\Phi}\tau_0^2 G_{MN}\partial_j X^N
\hat{\tbA}^{ji}_S\det \tbA_{ij}]+2\partial_i[N_0V^2 e^{-2\Phi}
\tau_0^2B_{MN}\hat{\tbA}^{ji}_A\det \tbA_{ij}]+\partial_i[N^ip_M] \ .  \nonumber \\
\end{eqnarray}
We will analyze these equations of motion for two possible
configurations. The first one corresponding to the tachyon vacuum
and the second one corresponding to the zero tension limit.
\subsection{Tachyon Vacuum Solution}
It is easy to see that $T=T_{min},p_T=0$ where $\frac{dV}{dT}(T_{min})=0, V(T_{min})=0$ is the
solutions of the equation of motion.
Further, the equation of motion (\ref{eqhamtau}) together with the constraints
$\mG_i$ implies that $\tau_0$ is constant.
 In this case the remaining equations of motion simplify
considerably
\begin{eqnarray}
\partial_0 A_i&=&-2N_0B_{MN}\partial_i X^N G^{MK}\Pi_K+2N_0
(\tbA_S)_{ij}\pi^j+\partial_i V_0+N^jF_{ji}  \ , \nonumber \\
\partial_0\pi^i&=&\partial_k[N^k\pi^i-N^i\pi^k] \ , \nonumber \\
\partial_0 X^M&=&\pb{X^M,H_T}=2N_0 G^{MN}\Pi_N+N^i\partial_i X^M  \ ,
\nonumber \\
\partial_0 p_M&=&N_0\Pi_K \partial_M G_{KL}
\Pi_L+2N_0\partial_M B_{KL}\partial_i X^L \pi^i G^{KN}\Pi_N-2
\partial_i[N_0B_{KM}\pi^i G^{KN}\Pi_N]-\nonumber \\
&-& N_0\pi^i\partial_iX^K\partial_M G_{KL}\partial_jX^L\pi^j+
2\partial_i[N_0\pi^i G_{MN}\partial_j X^N\pi^j]+\partial_i[N^ip_M]  \ . \nonumber \\
\end{eqnarray}
To proceed further we introduce following projector
\begin{equation}
\triangle^i_j=\delta^i_j-\frac{g_{jk}\pi^k \pi^i}{\pi^mg_{mn}\pi^n} \ , \quad
g_{mn}=G_{MN}\partial_m X^M\partial_n X^N
\end{equation}
that obeys the relation $\triangle^i_j\pi^j=0 $ and
$\triangle^i_{ \ k}\triangle^k_{ \ j}=\triangle^i_{ \ j}$. In other words
it is a projector on directions transverse to $\pi^i$.
Then we can write
\begin{eqnarray}\label{Nisplit}
N^i=\triangle^i_{ \ j}N^j+\frac{N^j g_{jk}\pi^k
\pi^i}{\pi^m g_{mn}\pi^n}
\equiv N_\bot^i+N_{II}\pi^i  \ , \nonumber \\
\end{eqnarray}
where by definition $ N^i_\bot g_{ij}\pi^j=0$. Before we proceed further
we should mention that $\pi^i$ has the physical dimension $[\pi^i]=L^{-(p+1)}$
where $L$ is some length scale. Then is convenient to introduce dimensionless $\tilde{\pi}^i$ when we
write $\pi^i=\tilde{\pi}^i\tau_p$.
Using this notation we can introduce following
 derivative
\begin{equation}
\pi^i\partial_i=\tau_p\partial_\sigma \ .
\end{equation}
Now we return to the equation of motion for $\pi^i$
where we use the split (\ref{Nisplit})
\begin{eqnarray}
\partial_0\pi^i=\partial_k N_\bot^k\pi^i+\tau_p\partial_\sigma N_{II}\pi^i
-\tau_p\partial_\sigma N^i_\bot -\tau_p\partial_\sigma N_{II}\pi^i
-\tau_p\partial_\sigma \pi^i N_{II} \ .
\nonumber \\
\end{eqnarray}
Our goal is to find solution of this equation  when $\pi^i$ are constants. In this case
the constraint $\mG_V$ is automatically obeyed while the equation above takes
the form
\begin{equation}
0=\partial_k N_\bot^k\pi^i
-\tau_p\partial_\sigma N^i_\bot
\end{equation}
that is obeyed for all $i$ on condition when $N_\bot^i=\mathrm{const}$ that without lost of generality can be taken to be equal to zero. This choice also simplifies
considerably the equation of motion for $X^M,p_M$
\footnote{Note that the equations of motion for $A_i$ determine the time evolution
of $A_i$ as functions of $p_M$ and $X^M$.For that reason we will not analyze it
explicitly.}
\begin{eqnarray}
\partial_0 X^M&=&2N_0 G^{MN}\Pi_N+\tau_p N_{II}\partial_\sigma X^M  \ ,
\nonumber \\
\partial_0 p_M&=&N_0\Pi_K \partial_M G^{KL}
\Pi_L+2\tau_p N_0\partial_M B_{KL}\partial_\sigma X^L  G^{KN}\Pi_N-2\tau_p
\partial_\sigma[N_0B_{KM} G^{KN}\Pi_N]\nonumber \\
&-&\tau_p N_0\partial_\sigma X^K\partial_M G_{KL}\partial_\sigma X^L+
2\tau_p\partial_\sigma[ N_0G_{MN}\partial_\sigma X^N]+\partial_\sigma[N_{II}p_M] \ .  \nonumber \\
\end{eqnarray}
Now we argue that given system of the equations of
motion possesses fundamental string solution. Note that the Nambu-Gotto
action for the  fundamental string in general background has the form
\begin{equation}\label{NGact}
S=-\tau_F\int d\tau d\sigma [\sqrt{-\det
\gamma_{\alpha\beta}}-B_{MN}
\partial_\tau Z^M \partial_\sigma Z^N] \ ,
\end{equation}
where
\begin{equation}
\gamma_{\alpha\beta}=G_{MN}\partial_\alpha Z^M\partial_\beta Z^N \ ,
\alpha,\beta=\tau,\sigma \ .
\end{equation}
It is easy to find the Hamiltonian from (\ref{NGact}).
Explicitly, from (\ref{NGact})  we find momenta conjugate to $Z^M$ as
\begin{equation}\label{KM}
K_M=\frac{\delta \mL_{NG}}{\delta \partial_\tau Z^M}= \tau_F
\frac{G_{MN}\partial_\tau
Z^N\gamma_{\sigma\sigma}-G_{MN}\partial_\sigma
Z^N\gamma_{\tau\sigma}}{\sqrt{-\det \gamma}}+\tau_F B_{MN}
\partial_\sigma Z^N
 \ .
 \end{equation}
Then the bare Hamiltonian $\bK$ is zero
\begin{equation}
\bK=\int d\sigma(K_M\partial_\tau Z^M-\mL_{NG})=0
\end{equation}
while using (\ref{KM}) we find  two primary constraints
\begin{eqnarray}
\mK_\sigma \equiv K_M\partial_\sigma Z^M\approx 0 \ , \quad
\mK_\tau\equiv \frac{1}{\tau_F}\Psi_M
G^{MN}\Psi_N+\tau_F
\gamma_{\sigma\sigma}\approx 0 \ , \nonumber \\
\end{eqnarray}
where we defined
\begin{equation}
\Psi_M=(K_M-\tau_F B_{MK}\partial_\sigma Z^K) \ .
\end{equation}
Then the extended Hamiltonian has the form
\begin{equation}\label{HEstr}
\bK_E=\int d\sigma (\lambda_\tau \mK_\tau+\lambda_\sigma \mK_\sigma) \ ,
\end{equation}
where $\lambda_\tau,\lambda_\sigma$ are dimensionless Lagrange multipliers
since $\mK_\tau,\mK_\sigma$ have the physical dimensions $L^{-2}$.
Using (\ref{HEstr}) we derive following  equations of motions for $Z^M,K_M$
\begin{eqnarray}
\partial_\tau Z^M&=&\pb{Z^M,\bK_E}=\frac{2}{\tau_F}\lambda_\tau G^{MN}\Psi_N
+\lambda_\sigma
\partial_\sigma Z^M \ , \nonumber \\
\partial_\tau K_M&=&\pb{K_M,\bK_E}=
\frac{\lambda_\tau}{\tau_F} \Psi_P
\partial_M G^{PN}\Psi_N\nonumber \\
&+&2\lambda_\tau\partial_M B_{NK}\partial_\sigma Z^K G^{NP}\Psi_P
-2\partial_\sigma[\lambda_\tau B_{NM}G^{NP}\Psi_P]-
\nonumber \\
&-&\lambda_\tau\tau_F \partial_M G_{KL}\partial_\sigma Z^K \partial_\sigma Z^L+
2\tau_F\partial_\sigma[\lambda_\tau G_{MN}\partial_\sigma Z^L] \ .
 \nonumber \\
\end{eqnarray}
Now we see that the non-BPS Dp-brane at the tachyon vacuum
with constant electric flux possesses fundamental string solution on condition when we
identify $Z^M$ with $X^M$ and $p_M=\frac{\tau_p}{\tau_F}K_M$ together with $N_{II}=\frac{1}{\tau_p}\lambda_\sigma,
N_0=\frac{1}{\tau_p}\lambda_\tau$.
 It is very interesting
that this solution does not depend on all world-volume coordinates
of non-BPS Dp-brane but it only depends on  $\sigma$, where  $\sigma$ is
defined by the orientation of the electric flux on the world-volume of non-BPS
Dp-brane at the tachyon vacuum. We mean that this is a natural result if we recognize
that it is believed that at the tachyon vacuum
the non-BPS Dp-brane disappears. Further,  note
that the localization of the electric flux on the world-volume
of non-BPS Dp-brane does not have  physical meaning when the full world-volume
diffeomorphism invariance is preserved.
\subsection{Zero Tension Solution}
Now we will discuss another interesting special solution corresponding to the case
when $\tau_0=0$.
Then $\tmH_0$ has simplifies
considerably and has the form
\begin{equation}
\tmH_0=\Pi_M G^{MN}\Pi_N+p_T^2+\pi^i
(\tbA_S)_{ij}\pi^j \ .
\end{equation}
Let us now  combine $T$
with $X^M$ into $\bZ^A=(X^M,T)$ so that
\begin{equation}\label{Hamzerotension}
\tmH_0=\bPi_A \bG^{AB}\bPi_B+\pi^i \gamma_{ij}\pi^j \ , \quad
\tmH_i=\bp_A\partial_i\bZ^A \ ,
\end{equation}
where
\begin{equation}
\bPi_M=\Pi_M   \ , \bPi_T=p_T \ ,
\gamma_{ij}=\partial_i \bZ^A \bG_{AB}\partial_j\bZ^B \ ,
\bG^{MN}=G^{MN} \ , \bG^{TT}=1 \ .
\end{equation}
Now we see that the Hamiltonian density for the tensionless
Dp-brane has almost the same form as the Hamiltonian density of the
non-BPS Dp-brane at the tachyon vacuum with exception that there is
an additional embedding  mode $T$. In other words tensionless non-BPS Dp-brane
propagates in the space-time with an additional dimensions. It is also
clear that this theory possesses fundamental string solution where again
this string propagates in the higher dimensional space-time. Finally we should
also stress one important point. Naively we could expect that the tachyon
condensation on tensionless non-BPS Dp-brane in the form of the kink solution
could lead to an emergence of tensionless D(p-1)-brane. However as follows from the
 form of the Hamiltonian constraint (\ref{Hamzerotension})
  we see that there is no tachyon potential
in the tensionless limit and hence it is not
possible to find the tachyon kink solution with the
interpretation as a lower dimensional tensionless D(p-1)-brane. This has also nice
physical interpretation since zero tension solution is very similar
to the tachyon vacuum solution where we argued unstable
Dp-brane should disappear and hence it does not make sense
to speak about lower dimensional tensionless D(p-1)-brane.
\subsection{Inclusion of RR fields: The Case of non-BPS D2-brane}
In this section we perform Hamiltonian analysis of non-BPS D2-brane with
the presence of the background Ramond-Ramond fields. We consider this specific
example  in order to simplify the form of $\star\tmG$ keeping in mind that the
generalization of this analysis to the more general case  is straightforward. Explicitly, in case
of non-zero RR fields $\star\tmG$ for unstable D2-brane has the form
\begin{eqnarray}
\star\tmG=\partial_\mu \omega^\mu-\epsilon^{\mu_1\mu_2\mu_3}
(\frac{1}{2}V(T)C_{\mu_1\mu_2}\partial_{\mu_3}T+\frac{1}{2}V(T)\mF_{\mu_1\mu_2}C\partial_{\mu_3}T) \ .
\nonumber \\
\end{eqnarray}
Now we proceed to the Hamiltonian formalism. The momenta conjugate to $v,A_0$ and
$\omega^i$ are primary constraints
\begin{equation}
p_v\approx 0 \ , \tau_i\approx 0 \ , \pi^0\approx 0
\end{equation}
while the momenta conjugate to $T$ $X^M$ and $A_i$ have the form
\begin{eqnarray}
p_T&=&\frac{e^{-2\Phi}V^2}{v}
\partial_\mu T (\tbA^{-1})^{\mu 0}_S\det \tbA-\frac{V}{v}(C_{12}+\mF_{12}C)\star \tmG
\nonumber \\
p_M&=&\frac{e^{-2\Phi}V^2}{v}\left(G_{MN}\partial_\mu X^N (\tbA^{-1})^{\mu 0}_S+B_{MN}
\partial_i X^N(\tbA^{-1})^{i 0}_A\right)\det \tbA \nonumber \\
&-&\frac{\star\tmG V}{v}\epsilon^{i_1 i_2}(C_{MN}+B_{MN}C)\partial_{i_1}X^N\partial_{i_2}T
 \ , \nonumber \\
\pi^i&=&\frac{e^{-2\Phi}V^2}{v}(\tbA^{-1})^{i0}_A \det \tbA-\frac{\star\mG V}{v}C
\epsilon^{ij}\partial_{j}T \ , \quad \tau_0=\frac{\star\tmG}{v} \ , \nonumber \\
\end{eqnarray}
where $\epsilon^{12}=-\epsilon^{21}=1 $.
Following the same procedure as in previous sections we find the extended  Hamiltonian in
the form
\begin{equation}
H_E=\int d^2\xi \left(\frac{ve^{2\Phi}}{2\det \tbA_{ij}V^2}
\tmH_0+N^i\tmH_i+\omega^i\partial_i\tau_0+N^i\tmH_i\right) \ ,
\end{equation}
where
\begin{equation}
\tmH_0=\Pi_M G^{MN}\Pi_N+\Pi_T^2+\Pi^i (\tbA_S)_{ij}\Pi^j
+V^2 \tau_0^2 e^{-2\Phi}\det \tbA_{ij}  \ ,
\end{equation}
where $\Pi_M,\Pi_T$ and $\Pi^j$ are defined as
\begin{eqnarray}\label{PiRR}
\Pi_T&=&p_T+
\tau_0 V (C_{12}+\mF_{12})\tau_0 \ , \nonumber \\
\Pi_M&=&p_M-B_{MN}\partial_i X^N\pi^i+\tau_0 V C_{MN}\partial_i X^N \epsilon^{ij}\partial_j T \ ,
\nonumber \\
\Pi^i&=&\pi^i+\tau_0 V C \epsilon^{ij}\partial_j T \ .\nonumber \\
\end{eqnarray}
We could analyze this system in the same way as in previous sections. However we see
from the form of the Hamiltonian density and from (\ref{PiRR})  that
non-BPS D2-brane at the tachyon vacuum does not couple to the Ramond-Ramond fields and
the dynamics of the configuration with the non-zero electric flux reduces
to the dynamics of the Nambu-Gotto string in this background. This is a non-trivial
result that supports the conjecture that the end point of the tachyon condensation
on the world-volume of non-BPS Dp-brane is the gas of the tensile strings.
\vskip .5in \noindent {\bf Acknowledgement:}
\\
This  work  was supported by the Grant Agency of the Czech Republic
under the grant P201/12/G028.

\end{document}